
\def\PRD{Phys. Rev. }\def\MPLA{Mod. Phys. Lett. }\def\NPB{Nucl. Phys. }
\def\CQG{Class. Quantum Grav. }
\def\ads{anti-de Sitter }
\def\bh{black hole }\def\st{spacetime }
\def\bg{background }
\def\Sch{Schwarzschild }\def\coo{coordinates }
\def\Mi{Minkowski }
\def\tran{transformations }
\def\sch{Schwarzschild }\def\GR{General Relativity}
\def\hyp{hyperboloid }\def\as{asymptotically }

\def\ds{ds^2=}\def\ef{e^{2(\phi-\phi_0)}}
\def\kp{{1+k\over 2}}\def\km{{1-k\over 2}}
\def\kf{{1-k\over 1+k}}\def\kk{{2k\over 1+k}}\def\kd{{-2\over 1+k}}
\def\kc{{2\over 1+k}}\def\R{{\cal R}}
\def\kg{{1+k\over 1-k}}\def\tpsi{{\tilde\psi}}
\def\lo{\lambda_1^2}\def\lt{\lambda_2^2}\def\efp{e^{2(\phi -{q\over 3}\psi)}}
\def\ha{{1\over 2}}\def\mo{{-1}}
\def\kl{{k-1\over k+1}}
\magnification=1200
\line {February 1994 \hfil  INFNCA-TH-94-4}
\vfill
\centerline {\bf DUALITIES OF LORENTZIAN AND EUCLIDEAN BLACK HOLES IN}
\centerline {\bf TWO-DIMENSIONAL STRING-GENERATED MODELS}
\vskip 1.2in
\centerline {\bf M. Cadoni}
\centerline {\bf S. Mignemi}
\vskip .2in

\centerline {Dipartimento di Scienze Fisiche, Universit\`a di Cagliari}
\centerline {and Istituto Nazionale di Fisica Nucleare, Sezione di Cagliari}
\centerline {Via Ada Negri 18, I-09127 Cagliari, Italy }\vfill
\centerline {\bf ABSTRACT}

We discuss the properties of lorentzian and euclidean \bh solutions
of a generalized 2-dim dilaton-gravity action containing a modulus field,
which arises from the compactification
of heterotic string models. The  duality symmetries of these solutions are
also investigated.
\eject

Two-dimensional dilaton-gravity theories have been studied for several
reasons. First of all, they are a theoretical laboratory for investigating
some features of gravity which are out of  our reach in four dimensions [1].
Moreover, they have great relevance in the study of 2-dim conformal field
theories and consequently of exact propagation of strings in fixed backgrounds
[2]. Finally, they arise naturally as effective 2-dimensional models of the
low-energy 4-dimensional string theory and are especially helpful in the study
of the scattering of radiation from the corresponding black hole solutions [3].

Studying an effective 4-dim string model, where non trivial couplings
of modulus fields arising from one-loop string threshold effects were taken
into
account [4],
we were lead to consider a 2-dim modulus-dilaton-gravity model [5]
whose essential feature is a 1-parameter action, which
contains as special cases most of the 2-dim models proposed till now.
Starting from a quite different approach, a model leading to equivalent
solutions was obtained in ref. [6].
In this letter, we wish to study some general features of the lorentzian
and euclidean \bh solutions of these theories and discuss the duality \tran
which connect some of them. We also comment about the significance of
such duality \tran from the point of view of string theories.
\bigskip
Let us consider the 2-dim action:
$$S={1\over 2\pi}\int\sqrt g \ d^2x\ e^{-2\phi}\left[R+4(\nabla\phi)^2-{2\over
3}(\nabla\psi)^2+\lo-\lt\ \efp\right],\eqno(1)$$
where $\phi$ is the dilaton field and $\psi$ a modulus field, $q$ being a
coupling constant.
This action was derived as an effective 2-dim action for heterotic string
models in the extremal \bh limit, when modulus fields are taken into account
[5].

It was shown in [5] that exact solutions can be obtained for any value of $q$,
provided that $e^\psi=const\times e^{3\phi/q}$. In this case the action (1)
simplifies to:
$$S={1\over2\pi}\int\sqrt g\ d^2x\ e^{-2\phi}\left[R+{8k\over
k-1}(\nabla\phi)^2
+\Lambda\right],\eqno(2)$$
where
$k={3-2q^2\over 3+2q^2}$,\ $-1\le k\le 1$ and $\Lambda=\lambda^2_2$.

In this form, the action admits several interesting special cases for given
values of $k$ [6]. The limit $k=-1$ (i.e. $q\to\infty$) is the string
theoretic case [2, 3]. For $k=0$, one obtains the Jackiw-Teitelboim theory [1],
while $k=1$ is the 2-dim equivalent of General Relativity [8, 9].

Black hole solutions of (2), or equivalently of (1), subject to the condition
$e^\psi=const\times e^{3\phi/q}$, can be easily found, for any value of $k$,
in the unitary gauge $ds^2=-e^{2\rho(\sigma)}dt^2+d\sigma^2$ [5, 6].
They are given by:
$$\ds-\sinh^2(\kappa\sigma)\cosh^{2k}(\kappa\sigma)dt^2+d\sigma^2,$$
$$\ef=\cosh^{k-1}(\kappa\sigma),\eqno(3)$$
where $\kappa={\sqrt{|\Lambda|\over 2(1-k)}}$.
Of course, in the limit $k\to -1$, the solution reduces to that found in [2,
10].

Another class of solutions of (2) is given by \ads\st with linear dilaton:
$$\ds-e^{2(k+1)\kappa\sigma}dt^2+d\sigma^2,$$
$$e^{2(\phi -\phi_0)}=e^{(k-1)\kappa\sigma},\eqno(4)$$
(flat space for $k=-1$). This is the quantum ground state of the theory [5].

Also of interest is the case in which the effective cosmological constant
$\Lambda$  is negative, even if this is not compatible with string theory.
In this case, the solutions are periodic in $\sigma$ and are given by:
$$\ds-\sin^2(\kappa\sigma)\cos^{2k}(\kappa\sigma)dt^2+d\sigma^2,$$
$$\ef=\cos^{k-1}(\kappa\sigma).\eqno(5)$$
As special cases, for $k=0, 1$, one gets 2-dim de Sitter \st.
\bigskip

It is interesting to notice the existence of a duality symmetry in the action
(1), which generalizes that found in [7] in the $k=-1$ case.

In fact, substituting for the Ricci scalar its value $R=-2e^{2\rho}(\rho ''+
\rho '^2)$ in
the unitary gauge, the action can be written:
$$S=\int d^2x e^{-2\phi+\rho}\left[-2(\rho''+\rho'^2)+4\phi'^2-4\kg
\tpsi'^2+\lo-\lt e^{2(\phi-\tpsi)}\right],\eqno(6)$$
where $\tpsi=q\psi/3$.
One can then readily check that the transformation:
$$\eqalign{\rho&\to k\rho-2(k+1)\tpsi,\cr
\phi&\to {k-1\over 2}\rho+\phi-(k+1)\tpsi,\cr
\tpsi&\to-k\tpsi+{k-1\over 2}\rho},\eqno(7)$$
leaves the action invariant, modulo a total derivative.\footnote{$^\dagger$}
{We remark that the reduced action (2) is of course invariant under the
transformations: $\rho\to k\rho-2(k+1)\phi$;\ $\phi\to {k-1\over
2}\rho-k\phi$.}

The effect of the transformations (7) on the \bh solutions (3) is to exchange
the hyperbolic sine and cosine everywhere:
$$\ds-\cosh^2(\kappa\sigma)\sinh^{2k}(\kappa\sigma)dt^2+d\sigma^2,$$
$$\ef=\sinh^{k-1}(\kappa\sigma).\eqno(8)$$
The solutions (4), instead, are self-dual, i.e. invariant under the
transformation (7).
Analogously, the compact solutions (5) are not affected by the duality
transformations, except for a shift of the coordinate $\sigma$, changing the
sine into cosine.
\bigskip

The properties of the Lorentzian black holes have been discussed in extent in
[5] and [6]. We give here a short review, add some remarks and discuss the
effect of the duality \tran on the solutions.

In order to study the black hole solutions, it is useful to write the metrics
in Schwarz- schild coordinates.
In such \coo it is in fact possible to continue the metrics besides the
horizon at $\sigma=0$ and to get a more immediate insight of their physical
properties.

The new \coo are defined so that the metric takes the form
$\ds -A (r)dt^2+A^{-1}(r) dr^2$.
The general solutions in this gauge are:
$$\ds(br^2-ar^\kk)dt^2+(br^2-ar^\kk)^{-1}dr^2,$$
$$e^{2\phi}=(\sqrt b r)^{\kl},\eqno(9)$$
where $b={(1+k)^2\Lambda\over 2(1-k)}$ and $a$ is an integration constant.

The curvature scalar is given by:
$$R=-2\left(b+{k(1-k)\over(1+k)^2}ar^\kd\right).\eqno(10)$$

For generic $k$ (i.e. $k\neq 0,\pm 1$), there are four different cases:
\smallskip
{\noindent a)\quad$\Lambda>0,\ a>0$:}

These are the solutions (3), obtained from the previous expression  by the
change of variables $r=(a/b)^\kp$ $\cosh^{1+k}(\kappa\sigma)$. They correspond
to asymptotically \ads black holes with horizon at $r_0=
(a/b)^{1+k\over 2}$ and singularity at $r=0$.
\smallskip
{\noindent b)\quad$\Lambda>0,\ a<0$:}

These are the duals of the previous solutions. They can be obtained from (9)
with the change of variables $r=(-a/b)^\kp\sinh^{1+k}(\kappa\sigma)$.
A naked singularity is present at finite distance ($r=0$).
\smallskip
{\noindent c)\quad$\Lambda>0,\ a=0$:}

This case corresponds to \ads \st (4). It is an everywhere regular spacetime.
\smallskip
{\noindent d)\quad$\Lambda<0,\ a>0:$}

For negative $\Lambda$, $a$ must be positive in order to have the correct
signature
for the metric. These spacetimes correspond to de-Sitter \bg with a naked
singularity at the origin and a cosmological horizon at $r_0=
\left({a\over b}\right)^{1+k\over 2}$. They can be
interpreted as the interior of the black holes (a).
\bigskip
The black holes solutions can be maximally extended by introducing Kruskal
coordinates.
In case (a) they can be obtained by defining a new coordinate (for simplicity
we set $|a|=|b|=1$):
$$r^*={1\over k+1}\int{dr\over r^2-r^\kk}={r^\kf\over k-1}\ {\rm F}\left({1-k
\over 2}, 1, {3-k\over 2}, r^{2\over 1+k}\right),\eqno(11)$$
where F is the hypergeometric function, and
$$u=e^{r^*+t},\qquad\qquad v=e^{r^*-t},\eqno(12)$$
giving rise to
$$ds^2=(k+1)^2\left(r^2-r^\kk\right)\exp\left[-{2\ r^\kf\over k-1}
{\rm F}\left({1-k\over 2}, 1, {3-k\over 2}, r^\kc\right)\right]dudv,\eqno(13)$$
where $r$ is defined implicitly in terms of $uv$ by (11) and (12).

In these coordinates, the metric is clearly regular at the horizon $uv=0$,
and in general singular at $uv=1$, i.e. $r=0$, excepted the special cases $k=0,
1$. Spatial infinity is represented by the curve $uv=-1$. The Kruskal
diagram for a generic
solution is shown in figure (1). It is quite similar to that of a \sch -\ads
\bh. The region I corresponds to the external spacetime, whose spatial
infinity is given  by the curve $uv=-1$. Region II corresponds to the \bh
interior. An observer in region II will fall in a finite proper time into the
singularity at $uv=1$. Regions III and IV have the usual white hole
interpretation.

In the $k=-1$ limit, as shown by Giveon [7], region V corresponds to the dual
\bh, with the naked singularity at $uv=1$. In fact, in that limit, the Kruskal
\coo have the same analytical form in both cases, but with the regions I and
V interchanged. This is a consequence of the dual symmetry [7]. We believe
that something analogous happens also in the more general cases. However, due
to the intricate analytical form of the metric (11) we were not able to check
this fact explicitly, except for $k=0$
\bigskip

One can classify the solutions according to the value of their masses and
of their thermodynamical parameters.
The mass can be easily obtained in cases (a) and (b) by means of the ADM
procedure. In the \sch gauge the result is [5]:
$$M={(1-k)b^{1-k\over2(1+k)}\over 2(1+k)}\ a.\eqno(14)$$
The sign of the mass coincides therefore with that of the constant $a$: the
dual solutions have negative mass. This is analogous to 4-dim GR, where naked
singularities correspond to negative masses.

The temperature is given as usual by the inverse periodicity of the regular
euclidean
solution and can be obtained by using the standard formula for the \sch\coo
[11]:
$$T={1\over 4\pi}{d\over dr}g_{00}\Big|_{r=r_0}={a^\kp b^\km
\over 2\pi(k+1)}
={1\over2\pi}\left({2M\over 1-k}\right)^\kp\left({\sqrt b\over 1+k}\right)^\km.
\eqno(15)$$
This is the temperature associated with the horizon in case (a). It is apparent
that it is a monotonically increasing function of the mass (at variance with
4-dim \GR) for any value of $k$, except the limit case $k=-1$, when it is
independent of $M$, $T={\sqrt b\over 4\pi}$. In the other limit case, $k=1$,
one has $T={M\over 4\pi}$.
In every case except $k=-1$, $M=0$ is therefore the endpoint of the
semiclassical  Hawking evaporation process.

The entropy can be most easily obtained by integrating the thermodynamical
relation $dS=T^{-1}dM$, resulting in:
$$S=2\pi\left(\kg\ {2M\over\sqrt b}\right)^\km.\eqno(16)$$
Again, we have a positive power function of M, with the exception of $k=1$,
when the entropy is given by $S=4\pi\ln M$ [8].
A remarkable relation, valid whenever $k\ne 1$ is $ST={2M\over 1-k}$.

It should be noticed that the formulae above are formally valid also for the
cosmological horizon in case (d).

It is also interesting to study the Newtonian limit of the \bh solutions (a).
If one neglects matter couplings to the dilaton, this is given as in GR by the
linearization of $g_{00}$. Besides the long-range force due to the \ads
asymptotic behaviour, one can have quite different short-range behaviour
depending on the sign of $k$: the short-range potential is in fact given
by $-ar^\kk$, it can therefore either diverge or go to zero at the origin.
Moreover, the short-range force is repulsive if $k>0$ (see figure 2). This
fact causes a very different behaviour in the quantum evaporation process [5].
\bigskip
We discuss now in some detail the case $k=0$.  This is the simplest non-trivial
example, but nevertheless possesses very interesting solutions, among which a
non-singular \bh [12] and can clarify some features of the more general case.
Its physical interest derives from its equivalence to the Jackiw-Teitelboim
theory [1].
The other special cases, $k=\pm1$, are discussed in detail in [2, 10] and [8].

As is easily seen, the solutions of the theory have constant curvature
$-\Lambda$ and the dilaton, in the \sch gauge, is proportional to $r^\mo$.
Contrary to the general case, there are no curvature singularities.
In two dimensions there exist several different manifolds with negative
constant curvature: the
solutions can indeed be divided in four classes:\footnote{$^\dagger$}{For
simplicity we put $|a|=|b|=1$ in the following discussion.}
\smallskip
{\noindent a) $\quad\Lambda>0$, $\ a>0$;
$\quad\ds-(r^2-1)dt^2+(r^2-1)^{-1}dr^2=-\sinh^2\sigma dt^2+d\sigma^2.$}

These solutions are non-singular black holes with a horizon at $r=1$ and \ads
asymptotic behaviour. They have been studied in ref [12]. The change of \coo
$$u=\sqrt{r-1\over r+1}\ e^t,\qquad\qquad v=-\sqrt{r-1\over r+1}\ e^{-t},$$
takes the metric into the Kruskal form:
$$\ds -4{dudv\over (1+uv)^2},\eqno(17.a)$$
with dilaton field:
$$\ef={1+uv\over 1-uv}.\eqno(17.b)$$
The Kruskal diagram is depicted in fig. 3.
The \Sch \coo cover the region I of the diagram (the region IV reversing
the signs of both $u$ and $v$). The curves $uv=0$
are of course the horizon $r=1$, while the hyperbola $uv=-1$ represents
spatial infinity.

The metric can be realized in 3-dim Minkowski space of signature $(-\ -\ +)$,
by considering a 2-sheet hyperboloid of equation $x^2+y^2-z^2=-1$ with the
choice of coordinates:
$$x=\sinh\sigma\sin t,\qquad\qquad y=\sinh\sigma\cos t,\qquad\qquad z=\cosh
\sigma.$$
The vertex of the \hyp corresponds to the horizon. This geometric
construction displays intuitively the regularity of the metric.
\smallskip
{\noindent b) $\quad\Lambda>0$, $\ a<0$;
$\quad\ds-(r^2+1)dt^2+(r^2+1)^\mo dr^2=-\cosh^2\sigma dt^2+d\sigma^2$.

This metric is dual to (a) and can be readily recognized to be the usual
2-dim \ads\st. It is an everywhere regular solution and can be realized
analogously to case  (a) in 3-dim \Mi space, but this time by a 1-sheet \hyp
of equation $x^2+y^2-z^2=1$ parametrized as:
$$x=\cosh\sigma\sin t,\qquad\qquad y=\cosh\sigma\cos t,\qquad\qquad z=\sinh
\sigma.$$
The Kruskal coordinates can be defined as:
$$u=\tan\left[\ha(t+\arctan r)\right],\qquad\qquad v=\tan\left[\ha(t-\arctan r)
\right],$$
so that the metric takes the same form as (17). The original coordinates,
however, cover in this case the region  V
of figure 3 (the region VI reversing
the signs of both $u$ and $v$), the hyperbola $uv=-1$
representing spatial infinity, and the dilaton field is given by:
$$\ef={1+uv\over u-v}.\eqno (18)$$
\smallskip
{\noindent c) $\quad\Lambda>0$, $\ a=0$;
$\quad\ds-r^2dt^2+r^{-2}dr^2=-e^{2\sigma}dt^2+d\sigma^2$.}

We finally consider the case where $a=0$. This is like (b) \ads spacetime,
but with a different parametrization. Now, the 1-sheet \hyp of (b) is in fact
parametrized by the \coo
$$x=e^\sigma t,\qquad\qquad y=\cosh\sigma-{1\over 2}e^\sigma t^2,\qquad\qquad
z=\sinh\sigma+{1\over 2}e^\sigma t^2,$$
which however only cover half of it. One can define Kruskal coordinates,
$$u={1\over r}+t,\qquad\qquad v=\left({1\over r}-t\right)^\mo,$$
such that the metric has again the form (17). In this case $u$ and $v$ can
take any real value and the metric is in fact self-dual.

\smallskip
{\noindent d) $\quad\Lambda<0$, $\ a<0$;
$\quad\ds-(1-r^2)dt^2+(1-r^2)^\mo dr^2=-\sin^2\sigma dt^2+d\sigma^2$.}

Case (d) is de Sitter spacetime. As is well known it can be represented by a
1-sheet hyperboloid of equation $x^2+y^2-z^2=1$, like \ads, but with the
signature of the embedding space changed to $(+,\ +,\ -)$. The metric is
regular, but a cosmological horizon is present at $r=1$. This spacetime can
be interpreted as the interior of the \bh (a).
In Kruskal coordinates,
$$u=\sqrt{1-r\over 1+r}e^t,\qquad\qquad v=\sqrt{1-r\over 1+r}e^{-t},$$
the form of the metric and of the dilaton is in fact again (17), and the range
of the \coo covers the region II of figure 3 (the region III reversing
the signs of both $u$ and $v$).
\bigskip
We pass now to consider the euclidean \bh solutions of our theory.
These are relevant in the sigma-model
formulation of 2-dim conformal field theories and have nice geometric
properties.
They can be simply obtained by analytical continuation of the lorentzian
solutions; for $\Lambda>0$:
$$\ds\sinh^2(\kappa\sigma)\cosh^{2k}(\kappa\sigma)d\tau^2+d\sigma^2,$$
$$\ef=\cosh^{k-1}(\kappa\sigma),\eqno(19)$$
or, for the dual case,
$$\ds\cosh^2(\kappa\sigma)\sinh^{2k}(\kappa\sigma)d\tau^2+d\sigma^2,$$
$$\ef=\sinh^{k-1}(\kappa\sigma),\eqno(20)$$
where $\tau$ is the euclidean time and $\kappa$ is defined as in eq. (3). All
the solutions but $k=-1$ tend asymptotically to euclidean \ads space, which is
also a solution of the field  equations:
$$\ds e^{2(k+1)\kappa\sigma}d\tau^2+d\sigma^2,$$
$$e^{2(\phi -\phi_0)}=e^{2(k-1)\kappa\sigma}.\eqno(21)$$
All the solutions (19) are regular for $0<\sigma<\infty$ and
$0\le\tau\le\beta$,
where $\beta=2\pi\sqrt{2(1-k)\over\Lambda}$ is the inverse temperature in this
gauge, and are therefore periodic in $\tau$. The dual solutions (20) are
instead singular at $\sigma=0$.
The corresponding geometries are depicted in figure 4.

In the case $k=-1$ the solutions have the well known "cigar" and "trumpet"
shape [2, 7] and are \as a flat cylinder. In the other cases, instead, the
solutions (19) have the shape of an infinite hood, which asymptotes a cone at
infinity. The dual solutions (20) are similar, but with a singularity
at finite
distance at $\sigma=0$, except the cases $k=0, 1$, when they are regular
everywhere and can be represented as hyperboloids in 3-dim Minkowski space.
The volume and the action of these solutions are of course infinite, so they
cannot be interpreted as instantons in the usual sense.

In the case $\Lambda<0$ the solutions are instead compact. They are periodic
both in $\sigma$ and $\tau$ and are given by:
$$\ds\sin^2(\kappa\sigma)\cos^{2k}(\kappa\sigma)d\tau^2+d\sigma^2,$$
$$\ef=\cos^{k-1}(\kappa\sigma).\eqno(22)$$
When $k=0, 1$ they are 2-spheres. In the other cases they are surfaces of
revolution, singular at $\sigma=0$. They have finite volume ${8\pi\Lambda
\over 1+k}$, but infinite action. Only for $k=-1$ the volume diverges, but the
action becomes finite.
\bigskip
To conclude, we add some remarks about the meaning of the duality symmetry of
the action (1) at
a string theoretic level. The issue one wants to clarify is the relationship
between the transformations (7) and the target space duality \tran of the
effective action of the string. A related, and even more fundamental  question
concerns the possibility that the background defined by (3) has a description
in terms of a 2-dim conformal field theory on the worldsheet. In ref. [5] the
action was derived as an effective 2-dim gravity action stemming from a 4-dim
effective string action in which string one-loop threshold effects were taken
into account.

It seems impossible to relate directly the symmetries of the 2-dim effective
theory to those of the original model from which it was derived. However,
some insight into the subject can be achieved by considering the 2-dim theory
as a toy model for 2-dim dilatonic gravity coupled to a modulus field.

We shall assume that the field $\psi$ appearing in the action (1) parametrizes
the radius $\R$ of some compactified manifold ($\R= e^{-\psi}$). In the large
radius limit,
which can be achieved by letting $q\to\infty$, the potential term for the field
$\psi$ disappears from the action:
$$S={1\over 2\pi}\int\sqrt g \ d^2x\ e^{-2\phi}\left[R+4(\nabla\phi)^2-{2\over
3}(\nabla\psi)^2+\lo\right].\eqno(23)$$

The ensuing equations of motion force the field $\psi$ to be constant, so that
the general solutions, in the unitary gauge, are:
$$e^{2\rho}=\tanh^2(\kappa\sigma),\qquad\ef=\cosh^{-2}(\kappa \sigma),
\qquad e^{-2\psi}=const.
\eqno(24)$$
The \st described by these solutions can be thought as the $D=3$ conformal \bg
of ref. [13], corresponding to the uncharged black string \st
$SL(2,R)/U(1)\times U(1)$, obtained by attaching a circle of constant radius to
every  point of the $D=2$ \bh \st of the $SL(2,R)/U(1)$ gauged WZW model.

The action (23) in the unitary gauge possesses two independent sets of duality
symmetries:
$$\psi\to -\psi,\eqno(25.a)$$
$$\rho\to-\rho, \qquad\phi\to -\rho+\phi.\eqno(25.b)$$
The \tran (25.a) corresponds to the well-known $\R\to 1/\R$ circle duality,
whereas the \tran (25.b) is the target space duality of ref. [7].

The former dualities are symmetries of the action (1) only in the large radius
limit. The exact duality symmetries of the action (1) are represented by eq.
(7), from which we recover in the limit $q\to\infty$ the form (25).

In conclusion, the introduction of the term $\lt e^{-2q\psi/3}$ in the action
(23) not only modifies radically the form of the solutions, allowing for
non-constant values of the field $\psi$, but also breaks the duality symmetries
(25). The general action (1) is, however, still invariant under the duality
transformation (7). It remains open the question whether the \bg defined by (3)
has an interpretation as a 2-dim conformal field theory. Work in progress
indicates that this is the case at least for $k=0$.
In fact in this case the solution describes the group invariant metric
of SL(2,R) and corresponds to the extremal limit of the charged black string
of ref. [13].

\bigskip

\centerline {\bf References}
\smallskip
\halign{#\quad&#\hfil\cr
[1]& C. Teitelboim, in {\sl Quantum Theory of gravity }, S.M. Christensen,
ed. (Adam Hilger,\cr & Bristol, 1984); R. Jackiw, {\sl ibidem};\cr
[2]&E. Witten, \PRD {\bf D44}, 314 (1991);\cr
[3]&C.G. Callan, S.B. Giddings, J.A. Harvey and A. Strominger, \PRD {\bf D45},
1005\cr &(1992);\cr
[4]&M. Cadoni and S. Mignemi, \PRD {\bf D48}, 5536 (1993);\cr
[5]&M. Cadoni and S. Mignemi, preprint INFN-CA-20-93, hep-th 9312171;\cr
[6]&J.P.L. Lemos and P.M. S\'a, Preprint DF/IST-9.93, to appear on \PRD D;\cr
[7]&A. Giveon, \MPLA {\bf A6}, 2843 (1991);\cr
[8]&R.B. Mann, A. Shiekh and I. Tarasov, \NPB {\bf B341}, 134 (1990);\cr
 &R.B. Mann, Gen. Rel. Grav. {\bf 24}, 433 (1992);\cr
 &R.B. Mann, T.G. Steele, \CQG {\bf 9}, 475 (1992);\cr
[9]&J.P.L. Lemos and P.M. S\'a, Preprint DF/IST-17.93, to appear in \CQG\cr
[10]&G. Mandal, A.M. Sengupta and S.R. Wadia, \MPLA  {\bf A6}, 1685 (1991);\cr
[11]&J.W. York, \PRD {\bf D31}, 775 (1985);\cr
[12]&J.P.L. Lemos and P.M. S\'a, Preprint DF/IST-8.93;\cr
[13]&S.H. Horne and G.T. Horowitz, \NPB {\bf B368}, 444 (1992).\cr}
\end